\documentclass[conference]{IEEEtran}
\IEEEoverridecommandlockouts
\usepackage{cite}
\usepackage{amsmath,amssymb,amsfonts}
\usepackage{algorithmic}
\usepackage{graphicx}
\usepackage{textcomp}
\usepackage{xcolor}
\usepackage{tabularx}
\usepackage{listings}
\usepackage{nopageno}
\usepackage{fancyhdr}
\def\BibTeX{{\rm B\kern-.05em{\sc i\kern-.025em b}\kern-.08em
    T\kern-.1667em\lower.7ex\hbox{E}\kern-.125emX}}

\begin{document}

\thispagestyle{empty}

\begin{huge}
IEEE Copyright Notice
\end{huge}

\vspace{5mm} 

\begin{large}
Copyright (c) 2020 IEEE
\end{large}

\vspace{5mm} 

\begin{large}
Personal use of this material is permitted. Permission from IEEE must be obtained for all other uses, in any current or future media, including reprinting/republishing this material for advertising or promotional purposes, creating new collective works, for resale or redistribution to servers or lists, or reuse of any copyrighted component of this work in other works.
\end{large}

\vspace{5mm} 

\begin{large}
\textbf{Accepted to be published in:} 

6th Annual Conf. on Computational Science \& Computational Intelligence (CSCI'19); Dec 05-07, 2019; Las Vegas, Nevada, USA;

https://american-cse.org/csci2019/\#!/home

\end{large}

\vspace{5mm} 

\begin{large}
DOI 10.1109/CSCI49370.2019.00032
\end{large}

\fancyhead[L,C]{}
\fancyhead[R]{2019 International Conference on Computational Science and Computational Intelligence (CSCI\textbf{)}}
\renewcommand{\headrulewidth}{0.4pt}

\newcolumntype{L}[1]{>{\raggedright\arraybackslash}p{#1}}
\newcolumntype{C}[1]{>{\centering\arraybackslash}p{#1}}
\newcolumntype{R}[1]{>{\raggedleft\arraybackslash}p{#1}}

\title{A Prevention and a Traction System for Ransomware Attacks}

\makeatletter
\newcommand{\linebreakand}{%
  \end{@IEEEauthorhalign}
  \hfill\mbox{}\par
  \mbox{}\hfill\begin{@IEEEauthorhalign}
}
\makeatother
\author{\IEEEauthorblockN{Murat Ozer}
\IEEEauthorblockA{\textit{School of Information Technology} \\
\textit{University of Cincinnati}\\
Cincinnati, Ohio, USA \\
ozermm@ucmail.uc.edu}
\and
\IEEEauthorblockN{Said Varlioglu}
\IEEEauthorblockA{\textit{School of Information Technology} \\
\textit{University of Cincinnati}\\
Cincinnati, Ohio, USA \\
varlioms@mail.uc.edu}
\linebreakand 
\IEEEauthorblockN{Bilal Gonen}
\IEEEauthorblockA{\textit{School of Information Technology} \\
\textit{University of Cincinnati}\\
Cincinnati, Ohio, USA \\
bilal.gonen@uc.edu}
\and
\IEEEauthorblockN{Mehmet F. Bastug}
\IEEEauthorblockA{\textit{Department of Interdisciplinary Studies} \\
\textit{Lakehead University}\\
Orillia, Ontario, CA \\
mbastug@lakeheadu.ca}
}

\maketitle

\begin{abstract}
Over the past three years, especially following WannaCry malware, ransomware has become one of the biggest concerns for private businesses, state, and local government agencies. According to Homeland Security statistics, 1.5 million ransomware attacks have occurred per year since 2016. Cybercriminals often use creative methods to inject their malware into the target machines and use sophisticated cryptographic techniques to hold hostage victims' files and programs unless a certain amount of equivalent Bitcoin is paid. The return to the cybercriminals is so high (estimated \$1 billion in 2019) without any cost because of the advanced anonymity provided by cryptocurrencies, especially Bitcoin \cite{Paquet-Clouston2019}. Given this context, this study first discusses the current state of ransomware, detection, and prevention systems. Second, we propose a global ransomware center to better manage our concerted efforts against cybercriminals. The policy implications of the proposed study are discussed in the conclusion section.\footnote{Preprint.  This paper was presented in 6th Annual Conference on Computational Science \& Computational Intelligence (CSCI'19); Dec 05-07, 2019; Las Vegas, Nevada, USA; https://americancse.org/events/csci2019} 
\end{abstract}


\section{Introduction}
Ransomware has been continuing to be a lucrative business for cybercriminals around the world. In 2019, many cities in the United States reported devastating ransomware attacks, as displayed in Figure \ref{fig3} \cite{GALLAGHER2019}. By 2021, experts estimate \$6 trillion ransomware related damages globally \cite{SteveMorgan2017}. In the US, cyberattacks are the fastest growing crime, which becomes more sophisticated and costly to victims, including big companies. The history of ransomware attacks begins with the invention of the Web in 1989. The first attack spread via floppy disk, which was requesting \$189 sent to a Panama post office. 
FBI arrested Ohio native perpetrator of the attack in a relatively short time because the form of the attack was simple, little or no anonymity was available, and less people had access to the internet at that time. 
Today, we have more than 1.2 billion websites with 3.8 billion internet users, and it is expected that we will reach 6 billion internet users by 2022. 

\begin{figure}
    \centering
    \includegraphics[width=1\columnwidth]{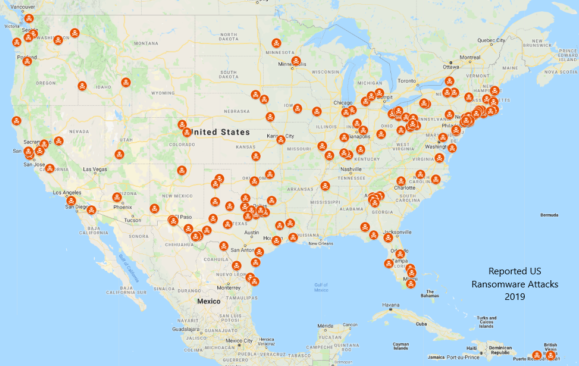}
    \caption{A Map of reported US ransomware attacks in 2019 by Armor\cite{armor}}
    \label{fig3}
\end{figure}

In addition to the websites, the  number of Internet of Things (IoT) devices has been increasing  exponentially and expected to reach 200 billion by 2020 \cite{Hung2017}. Putting it differently, today, cybercriminals may find more security deficits through wirelessly connected devices, which may include, but not limited to, smart home systems, pacemakers, and brain neurostimulators. 

Moreover, recent cyberattacks such as WannaCry and NotPetya revealed that criminals use more sophisticated techniques and encryption algorithms, which make it impossible to revert the data into its original format unless paying criminals via a cryptocurrency (mostly Bitcoin) and receiving a decryption key \cite{Symantec2019}.

The ransomware process starts with injecting malware into the networked computer by targeting human or technical weaknesses. Human related weaknesses usually stem from opening and clicking spam messages, which are also called phishing emails. Technical weaknesses may contain many different factors ranging from using publicly accessible Wi-Fi networks to a lack of firewall protection. After the infection process, as seen in Figure \ref{fig1}, cybercriminals change the file system by encrypting the whole computer files and allow a victim to see only their message and Bitcoin payment process. When cybercriminals hack a computer, it is nearly impossible to decrypt the files unless having the decryption algorithm or a decryption key. For this reason, victims tend to pay to cybercriminals to restore their hostage data from the criminals \cite{Duncan2019}.

There is a high debate not to pay cybercriminals for their ransomware attacks \cite{Symantec2019}; however, the end results of the cyber-attack can be much more devastating than the ransom amount due to the time-sensitive data such as health care data. For this reason, even FBI sometimes suggests paying back to criminals to restore the data \cite{dudley2019}. For instance, Baltimore city had fallen victim to a ransomware attack in 2019 and residents were even not able to perform basic payments. For this reason, Baltimore officials voted to transfer \$6 million worth of Bitcoin to cybercriminals to back to normality \cite{Duncan2019}. However, paying the ransom amount does not always guarantee to have the data back because the success rate of getting the decryption key from the cybercriminals is only 47\%. 

Furthermore, even though a decryption key is provided as a result of the payment, not all files can be completely restored as experienced in the Baltimore case. For this reason, ransomware detection, and preventive software and techniques are vastly discussed in the literature.

\begin{figure}
    \centering
    \includegraphics[width=0.6\columnwidth]{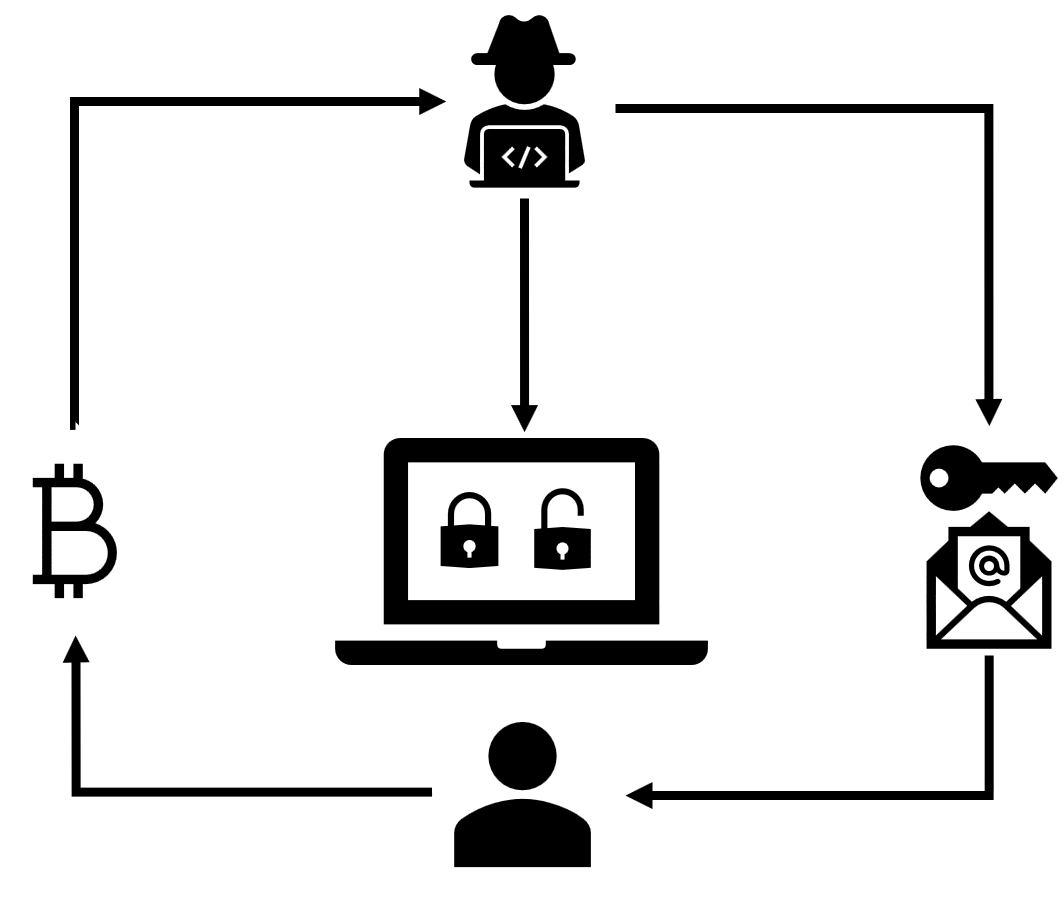}
    \caption{How Ransomware Works}
    \label{fig1}
\end{figure}

\section{Related Works}
Researchers have developed different systems to identify and prevent ransomware attacks. For instance, Scaife et al. \cite{Scaife2016} developed the CryptoDrop system, which basically scans the file system against suspicious activity. If CryptoDrop detects an abnormality, the software triggers and stops ransomware from executing the rest of the malware code to encrypt the files. In this way, the researchers state that their system can stop ransomware as little as losing 10 files out of 5100. 

Similarly, Kharraz et al. \cite{Kharraz2015} analyzed the anatomy of the ransomware attacks and discovered that even sophisticated ransomware attacks could be stopped after detecting abnormality in the file system activity. 

Continella et al. \cite{Continella2016} developed the ShieldFS system, which analyzes low-level filesystem activity to create a benchmarking filesystem from many different benign applications (like analyzing billions of low-level file activity). If the ShieldFS detects a malicious activity compared to benign ones, it immediately recovers all of the original files. 

Kolodenker et al. \cite{Kolodenker2017a} proposed a different light-weight system called PayBreak. This method created a symmetric session to imitate ransomware encryption keys so that victims can later use these copied symmetric keys to restore their files back without paying to the attacker. Further analysis showed that PayBreak could successfully decrypt files from twelve ransomware families. There are 30 known ransomware families. The authors assert that PayBreak adds negligible performance overhead to the computer processor, which in turn makes it preferable to run it all the time. 

Gomez-Hernandez et al. \cite{Gomez-Hernandez2018} offered a practical and easy solution (called R-Locker) by deploying a set of honeyfiles to the target environment. In this way, when the ransomware reads one of these files, it is automatically blocked. 

\section{Current Study}

The methods and software discussed in these studies usually require additional tools installed on the target system of users. However, most users do not install these tools on their computers unless they come as built-in. Even though detection systems are provided at the beginning, users fail to update the security patches on time, which again opens back doors to cybercriminals. For this reason, in addition to the advanced detection systems, we need to increase user-level ransomware awareness by promoting training against malicious activities such as phishing emails and visiting non-secure websites. On the other hand, more measures are needed at the agency level including strong firewall, file back-up system, email filtering service, two-factor authentication, and on-time system updates. 

Our study discusses that although detection and prevention systems harden the system vulnerabilities against ransomware attacks, it is nearly impossible to completely prevent these attacks because of the never-ending fight between good and bad actors, as is the case in every field of the life. For this reason, in addition to the detection and prevention systems, we need to create and maintain a concerted effort that tracks ransomware attacks to its originated person/group and location. As discussed above, a fair number of victims, including government organizations, tend to pay the ransomware. However, we know little about many of these victims since more than 50\% of ransomware targets small to mid-size governmental agencies or companies which, most of the time, stay private unless a data breach policy forces them to declare it to the public \cite{Duncan2019}.

Given this context, this study discusses a joint effort to track Bitcoin transactions until cashed-out Bitcoin addresses. Cybercriminals prefer Bitcoin transactions because it provides anonymity which makes nearly impossible to identify the identity of the target person. In addition to this, Bitcoin transactions are irrefutable, which basically guarantee not canceling or reversing the payment. Even though Bitcoin transactions maintain anonymity for both a sender and a receiver, it normally provides a pseudonym rather than anonymity thanks to the public ledger of the Blockchain. That means further analysis may identify the person that hides behind the walls of Bitcoin for their ransomware attacks. Tracing Bitcoin transactions back to the originated person/group or location is not easy but not impossible, as well. 

Therefore, the proposed study offers two techniques to uncover Bitcoin transactions. First, there are many empirical studies in the information technology area that discuss techniques to reduce the anonymity of bitcoin transactions \cite{Dupont2015}, \cite{Fleder2015}, \cite{Kumar2017}, \cite{Meiklejohn2016}, \cite{Nick2015}, \cite{Reid2013}, \cite{Wijaya2016}, \cite{Huang2018}, \cite{Paquet-Clouston2019}.

We will test these techniques against the public ledger of bitcoin transactions to unravel the anonymous nature of the bitcoin public keys. Each proposed technique from different peer-reviewed journals will be evaluated, and promising results will be pooled in a combined technique to de-anonymize ransomware related bitcoin transactions. Even though the complete de-anonymization might not be possible in certain cases, we will identify private keys by using a confidence interval. Thus, each transaction will have a score that reflects that confidence interval. The practical implementation of using confidence intervals will help the classification of the identified data into strata that facilitate evidence identification (e.g., a transaction belongs to a certain person with 90\% confidence).

Second, our model proposes using open sources to collect intelligence towards the identification of private bitcoin addresses, using sophisticated analytical tools (e.g., one-to-many mapping, link/graph analysis, clustering/pattern analysis, because criminals often leave digital footprints that help to detect their identity). For instance, in dark web forums, a person sends a message to a buyer to request the money (cryptocurrency) to his bitcoin address. We will aggregate this information and use pattern and keyword matching to identify the account holder's name, IP address, and location using various open-source data including social media. Once the bitcoin address is attached to a person, we will house that information in a separate database that holds blacklisted bitcoin addresses. As displayed in Figure \ref{fig2}, our model consists of ten steps:

\begin{enumerate}
\item Method-1: Test proposed algorithms by researchers to de-anonymize Bitcoin transactions used in ransomware attacks
\item Method-2: Identify ransomware related Bitcoin addresses and track the incoming and outgoing traffic as explained in Paquet-Clouston et al.'s study \cite{Paquet-Clouston2019}.
\item Method-3: Analyze for abnormalities in Bitcoin prices in order to detect likely ransomware transactions as explained in Jareth's study \cite{Jareth2019}
\item Method-4: Automatically check real-time transactions to capture Bitcoin addresses used in previous ransomware attacks
\item Server-1: Monitor Bitcoin Ledger
\item Server-2: Blacklisted Bitcoin data analysis
\item Action-1: Open Source analysis through automated scans to find pseudonyms of ransomware related Bitcoin addresses
\item Action-2: Analytical analysis (link analysis) that constantly augments  de-anonymization by associating previously unlinked Bitcoin addresses
\item Action-3: Manual data entry from different users/victims to increase de-anonymization process
\item Action-4: Constant feedback from the field to evolve the accuracy of the system
\end{enumerate}

\begin{figure}
    \centering
    \includegraphics[width=0.8\columnwidth]{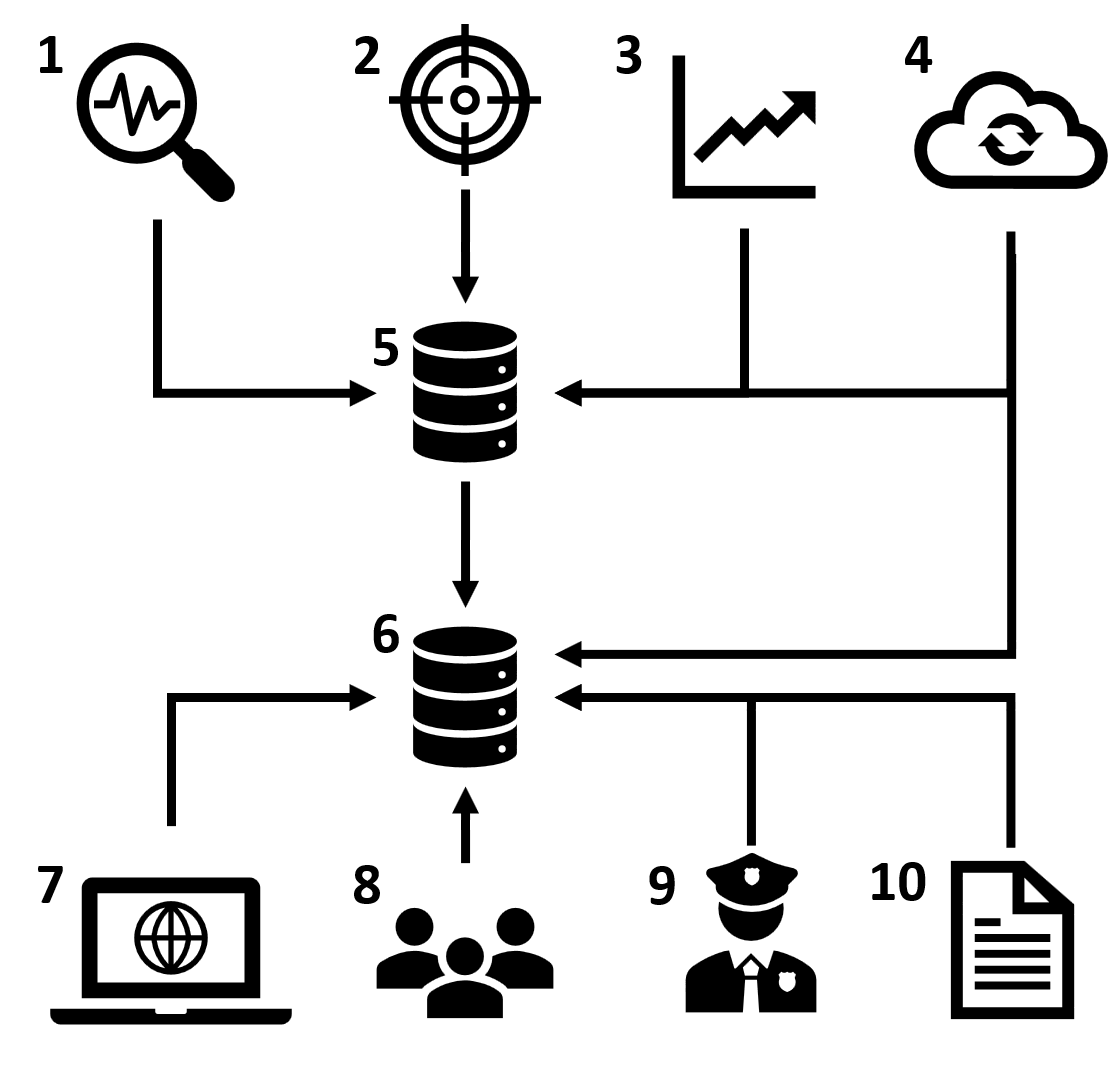}
    \caption{Model Workflow}
    \label{fig2}
\end{figure}

In addition to this, there are many websites that publish the wallet IDs of bitcoin addresses. For instance, www.walletexplorer.com shares wallet IDs even for dark web websites. This is a wealth of information that provides leverage for researchers and investigators to detect criminal cryptocurrency transactions \cite{WalletExplorer2019}. 

We will download disparate blacklisted bitcoin transactions into a platform (Server 2 in Figure 3) to build an algorithm that will link unconnected bitcoin transactions to one another in order to better understand these relationships in a mapped network. For this purpose, we will employ the principles of graph/link analysis that easily link related bitcoin addresses to each other as well as dark web websites \cite{Fleder2015}. The next step will be de-anonymizing blacklisted bitcoin addresses (e.g., who the person is, where the person is located) with open-source intelligence analysis, as explained above. The reason for using link analysis is that once we identify a chain/node in the network, we can better predict the other related unidentified bitcoin addresses. Putting it differently, when we identify a node and its related links, the anonymity of the Bitcoin addresses diminishes towards pseudonyms (i.e., wallet-id) and later to real identities through using open sources because even hackers or related associates leave some footprints on the internet \cite{Huang2018}. Using a similar strategy proposed in the current study, US Department of Justice was able to track and indict two Iranian men responsible for the SamSam ransomware attacks in different cities of Atlanta, Newark, New Jersey, the Port of San Diego, and Hollywood Presbyterian Medical Center in Los Angeles \cite{dudley2019}

Finally, our study offers to have a global center to better collect ransomware information, which helps to prevent better strategies. As discussed above, not all ransomware attacks are reported to FBI; therefore, we are not able to collect all ransomware cases which contain valuable information such as infection style, deployed encryption family of the ransomware, and Bitcoin address(es). In addition to this, as Dudley and Kao \cite{dudley2019} mentioned, there are many known cases that victims hired data recovery companies; however, those companies just paid back to cybercriminals to restore the data and never told the victims and the law enforcement officials about the case. For this reason, as the literature discusses to train users against cyber-attacks, this study also encourages to train victims to immediately share their cyber-attacks with the proposed ransomware center that collects the timely and necessary information to cope with the cybercriminals. 

\section{Conclusion}

The ransomware epidemic continues to spread due to its lucrative business. Increased digital activities, internet users, and the Internet of Things (IoT) devices create more vulnerabilities to cybercriminals, which require advanced detection, prevention, and tracking systems to deter likely cybercriminals. This study discussed that although researchers developed effective detection and prevention systems, many of them require installing those tools to the target machines, which always remain the possibility that some computers and systems may not have the required tools installed or updated. In addition to this, as cybersecurity personnel updates themselves against cyber-attacks, cybercriminals also try to invent new techniques to open back doors for their criminal activities. 

This study discusses that the realistic approach to ransomware attacks is that we will always experience a certain level of attacks regardless of having advanced detection and prevention systems. For this reason, in addition to the detection and prevention systems, this study proposes to create a global ransomware center that collects as much as possible data to develop adequate methods against cybercriminals and to identify individuals/groups behind these attacks. 

In the United States, FBI tries to play a central role in investigating ransomware attacks. However, the FBI usually keeps the attack details private and mostly does not share with researchers to conduct further analyses to develop better coping strategies against ransomware attacks. Our study proposes that academia, researchers, and law enforcement should work together around a global center because recent ransomware statistics suggests that ransomware is a global issue and threats/harms wide range of victim profiles ranging from daily internet users to sick and injured people waiting services at health-care agencies. For this reason, rather than mere law enforcement efforts, concerted efforts are needed around a global ransomware prevention and traction center to truly cope with ransomware attacks/threats which we are currently experiencing and will continue to experience in the future.

\bibliographystyle{IEEEtran}
\bibliography{Ransomware}

\end{document}